\documentclass[english]{article}
\usepackage{dirtytalk}
\usepackage[utf8]{inputenc}
\usepackage[english]{babel}
\usepackage[square,numbers]{natbib}
\usepackage{comment}
\usepackage[center,small,it]{caption}
\usepackage{tabu}
\usepackage{graphicx}
\usepackage{mathtools}
\usepackage{amssymb}
\usepackage{xspace}
\usepackage{algorithm}
\usepackage{algpseudocode}
\usepackage[hyphens]{url}
\usepackage{hyperref}
\hypersetup{colorlinks=true,breaklinks=true}
\usepackage{multicol}
\begin{document}

\title{A Certain Tendency Of The Database Community\footnote{Our title takes inspiration from Francois Truffaut's article ``A Certain Tendency Of The French Cinema'' from the \textit{Cahiers du Cin\'ema} in 1954.  In his case, the certain tendency was psychological realism; in ours, it is the single system image.}}
\author{Christopher S. Meiklejohn \\ Universit\'e catholique de Louvain \\ christopher.meiklejohn@gmail.com}
\maketitle

\begin{abstract}
We posit that striving for distributed systems that provide ``single system image'' semantics is fundamentally flawed and at odds with how systems operate in the physical world.  We realize the database as an optimization of this system: a required, essential optimization in practice that facilitates central data placement and ease of access to participants in a system.  We motivate a new model of computation that is designed to address the problems of computation over ``eventually consistent'' information in a large-scale distributed system.
\end{abstract}

\section{Eventual Consistency}

When we think about the physical world we live in, we do not usually say it is \textbf{eventually consistent} \cite{vogels2009eventually}, this is a term usually applied to computing systems, made up of multiple machines, that have to operate with ``shared'' information.

Eventual consistency is a consistency model for replicated, shared state.  A \textbf{consistency model} is a promise or contract between an application developer and a system that an application will run on.  Such a \textbf{contract} states the following: given the developer follows the rules defined by the system, certain outcomes from the system are guaranteed.  This makes it possible for developers to build successful applications, for without this contract, applications would have no guarantee that the actions they perform would have a correct outcome.

A \textbf{consistency model} specifies rules about when and where the effect of an update will be visible to the members in a distributed system that receive ``interactions'' in the form of protocol messages: or more formally, a consistency model defines possible partial orders and interleavings that events will be observed by different parties in the same system.  For instance, a \textbf{strong} consistency model such as linearizability \cite{herlihy1990linearizability}, states that as soon as you perform an update to a replica you can see the effects of that update immediately, or that the system closely approximates the physical time order of events.  A \textbf{weak} consistency model might place no bound on when the effect of an update will be visible.  \textbf{Eventual consistency} is a extremely weak consistency model and is only informally specified as follows:
\\

\say{...the storage system guarantees that if no new updates are made to the [replicated, shared] object, eventually all accesses [to any replica] will return the last updated value.} \cite{vogels2009eventually}
\\

Rather unfortunately, given this model is implemented by many real world systems, this statement does not say much about when a particular update may be visible to other members in the system.  This statement only states that at the end of a given execution, the most recent effect will be present at each replica of the object.

So, the question naturally follows: why are all of these consistency models required and why would one choose a rather weak model when you can pick a stronger model?

The answer is that \textbf{consistency}, or what guarantees we can make about when and where updates will be visible in the system, is fundamentally at odds with \textbf{availability}, the ability for a system to keep functioning in the event that some or all of the nodes in the system can not communicate with one another \cite{gilbert2002brewer}.  Weaker consistency models that offer fewer guarantees about when the effect of an update might be observed by other members in the system, can offer higher availability, allowing the system to continue to operate while components of the system are offline, where stronger models might prohibit updates when nodes can not communicate, to enforce that all nodes see all effects in the same order.

To demonstrate, let us take an example from Sebastian Burckhardt's work \cite{export:230852} on eventual consistency.  Consider the case where two machines are used to manage seat reservations for an airplane flight: we use two machines to increase our \textbf{tolerance to failure}, requiring that each machine must have a replica of all of the information.  Two users of the system, Tom and Chris, want to book the one remaining seat on the same flight.  Now, both Tom and Chris concurrently issue a request to different replicas.

If network connectivity between the two systems is interrupted at the time of the concurrent requests, the system must make one of two possible choices:
\begin{enumerate}
  \item The system can make a change to its local copy of the information and allow the concurrent seat requests to be executed.  However, this will result in conflicting information when connectivity is restored and the nodes exchange information about what happened during the period without connectivity.  How do we choose which update wins, and subsequently who no longer has a flight?\footnote{One trivial solution here is to arbitrate based on the temporal time when the update was performed, preferring the most recent update.  However, this strategy is problematic in many cases as it can lose valid updates.}
  \item The system can become \textbf{unavailable} and refuse to execute the seat requests when the network fails.  This attempts to treat distributed state across a set of replicas as a \textbf{single system image.}
\end{enumerate}

If we choose the \textbf{weak} consistency model that affords us higher availability, we need to address the problem of concurrent modifications to shared state.  In our previous example, how do we decide to resolve the concurrent seat bookings: who ``wins'' the seat?

What ever method we choose, the decision must be \textbf{deterministic} if we want to ensure that all copies of the shared state arrive at the same value once all of the updates have been delivered in the system.  Ideally these \textbf{conflict resolution} functions \cite{terry1995managing} should be tolerant to message reordering and message duplication \cite{shapiro2011comprehensive} to ensure correct execution on an unreliable, asynchronous network.

\section{The World is Eventually Consistent}
There is a reason that eventual consistency keeps coming up: it is fundamentally part of how systems in the real world operate.  Let us further explore this with a few analogies.  

\subsection{Recorded Knowledge}
\label{recorded_knowledge}
The first example looks at recorded knowledge in a particular field as an approximation of that knowledge in the physical world.

For instance, if the leading researchers on breast cancer were to document the state-of-the-art in a book, as the document is being written it would no longer reflect the state-of-the-art.  The collective knowledge of this group is always changing, and as long as we continue to rewrite the document it will only be approaching the combined knowledge of the group.  We can express this somewhat formally: if we had a way to view the group's knowledge as an omniscient observer and we represent that knowledge as a linear function, the recorded text would be asymptotic to the function of the sum of global knowledge.

\subsection{Message Passing}
Members of the same system exchange information by ``interacting'', or sending messages containing information to each other.  These messages between members of the system can be arbitrarily dropped and delayed, just like in traditional, unreliable, asynchronous networks.  We can find many analogs to these messages in the physical world: letters sent via the postal service, text messages, and telephone calls are all such examples.

Returning to the example of recorded knowledge in Section~\ref{recorded_knowledge}, we can think of the book itself as a member of the system.  This book would then receive messages periodically from the authors containing updates to existing information in the book, or new information that should be added to the book\footnote{We acknowledge that the idea of common knowledge in a distributed system can only be established \textit{a priori} \cite{halpern1990knowledge} and that knowledge can only grow asymptotic to common knowledge monotonically to the number of messages exchanged between members of the system.}.

\subsection{Primary Site}
We can think of each researcher as the \textbf{primary site} of their information.  They can exchange this information with others through message passing; but updates to this information are always coordinated through the primary site.  When we compile this information from a number of researchers into the recorded text, we have \textbf{composed} this information together.  This composition, in itself, can also be \textbf{replicated} between different parties.  

Consider the case where Tom and Chris are going to meet at a restaurant.  Tom and Chris both leave from different locations and begin to make their way to the restaurant.  Both have a general idea where the restaurant is located, but may not know the precise location.  In this example, Tom and Chris both make their way to the neighborhood where the restaurant is located, but then must consult a map on their smartphone to identify how to travel the remainder of the distance.

In this example, both Tom and Chris operated with multiple replicas of the information they needed to reach the restaurant:
\begin{itemize}
\item \textbf{Local replica.} Local information, stale and/or incomplete, that they both had in their heads;
\item \textbf{Stale replica.} Cached copy of a USGS map on their smartphone;
\item \textbf{Primary site.} Consulting the USGS for a map of the area from the USGS website.
\end{itemize}

%

\section{The Database as an Optimization}
We consider the idea of a planet or interplanetary-scale distributed system as a single connected component, where each node is a member of this system, and the primary site for some information.  The traditional \textbf{database} is an optimization of this graph.  When you take a subgraph and reduce it to a single node, or perform edge contractions for all edges in a particular subgraph, what results is an \textbf{overlay network}, or single connected component, consisting of \textbf{databases}.

Consider the following example.  If we take all of the information known by all of the people living in Japan and reduce it to a single node; similarly, we can do the same action for all of the United States, France, and the other countries in the world each as single nodes.  If we then connect each of these nodes together, the result resembles an \textbf{eventually consistent, partially-replicated, geo-replicated database}.

Designs such as this are very familiar in practice.  For instance, Facebook, a large social-network application on the web, has a single user profile for each user that is active in the system.  Each of these profiles is replicated across several of their data centers for performance; however, only one data center is deemed the primary site where all updates are performed.  Other social media sites, such as Twitter, also follow a similar design pattern.

If we return to our example of breast cancer research, we can imagine that researchers around the world keep the Wikipedia entry for this topic updated.  Referring to Section~\ref{recorded_knowledge}, this recorded knowledge on Wikipedia is just an \textbf{optimization} in the same way, given Wikipedia is just a replicated database of information on various subjects.

While just an optimization, this optimization is extremely important for several reasons.  It would be extremely impractical to have to hear information directly from the \textbf{primary site} every time you needed information: this optimization makes for extremely efficient distribution of information.  Therefore, this optimization is important for both providing low latency, high-availability, and fault-tolerance of the system.

We have designed many of our abstractions in computing with this optimization in mind: \textbf{clients} communicate with \textbf{servers} that are assumed to be the canonical source of some information.  Clients repeatedly retry sending state to a server where the information is not considered \textbf{persisted} unless the operation completes, when in reality the file living on the client is the source of truth.

As we enable Internet of Things applications and mobile applications running on smartphones become more and more commonplace, we continue to move closer to a design space where our traditional models of computation and centrally locating data becomes untenable.  Mobile applications treat the mobile device as a client in a system where a central database is used for storing and retrieving information; these designs implicitly assume that the central database is always reachable, which is often not true in areas of limited connectivity. These designs are inherently limiting: we impose a structure of a single primary site for information because it makes processing easier with the tools that we have today.

\section{The Living Database}
\textbf{We are moving towards large-scale edge computation.}  Given users are the primary site for their own information, the database challenge is largely a \textbf{constraint satisfaction problem}: how do we know where to send requests for information, where we supply some notion of how stale we allow that information to be, and how fast we need the information to be provided to us.  Given a long enough amount of time, we can wait for the canonical source of some information to respond to us; however, sometimes we need information faster to make some decision, and we will accept stale information.  If we contact a stale source today to service a request within some bounded amount of time, how can we determine that a source that we later contact has newer (or even more stale) information than ourselves\footnote{To be fair, even in these single data center designs, we still have to reason about stale information because we replicate data within the data center for higher availability and fault-tolerance.}?

Suppose we had \textbf{mergeable} data items that existed in time and space: with two copies of a set of information you could determine what copy they branched from, which was newer, and always combine into a single copy that resulted in the most up-to-date information between the two.  Given these items, we could share information between members of a system, and periodically ``refresh'' our values by merging in copies we receive from the other members on a network.

We can use these mergeable data structures as the basis for a programming model.  Consider the extension to a functional programming language where the primitive data structures in this language are all mergeable data structures.  In this model, the results of the executions of programs are also \textbf{mergeable}.  These computations carry along with them the information about the data items that have contributed to generating their result, and these computations themselves are also mergeable.  In systems like Spark, this may remind you of the notion of \textbf{lineage}.

In this model, each member of the system has some \textbf{partially-replicated knowledge} and some knowledge that they are the \textbf{primary site} for.  This information is exchanged between members of the system and merged with each member's local information: this provides both fault-tolerance, and lower latency in servicing requests for information from peers.

This knowledge provides the basis for local computations: for instance, using some local information to compute some result.  This result will approach the true global value asymptotically, but will never be reached.  The results of these computations will be \textbf{mergeable} themselves: for instance, if two members have computed over a subset of information.  This has an important analog in the real world: sometimes we care about the result of a computation without actually caring about some, or all, of the input required to generate that result.  \textbf{The results of computations are first class in this system.}

In this model, mobile clients and the servers that service them are \textbf{all equal.}  Clients exchange information and cache results to provide lower latency on servicing requests.  Clients are the source of truth for their own information. \textbf{Servers} are just clients that are located in a particular position in the network, and that are not the source of truth for any data; these nodes serve only to reduce latency in the system by replicating information.

\section{Conclusion}
So, when it has been said that Uber, a startup focused on ride sharing, ``[Uber] Goes Unconventional'' \cite{uber} by putting the canonical state regarding an in-progress ride on the drivers mobile devices, eschewing a traditional data center centric approach, how unconventional is it?  Isn't the phone the \textbf{primary site}?

\textbf{It is.}  As we continue to increase the number of globally connected devices, we must embrace a design that considers every single member in the system as the primary site for the data that it is generates.  It is completely impractical that we can look at a single, or a small number, of globally distributed data centers as the primary site for all global information that we desire to perform computations with.

In emerging countries, there still exist a number of devices that transmit data between members using USB devices, sneakernet techniques, and peer-to-peer communication.  It will most likely be a long time before these devices are connected directly via the public Internet, and we must provide computation models that allow for connected devices to communicate with each other and compute a result over some notion of shared data.

Can we build abstractions that allow devices to communicate peer-to-peer, acknowledging the true primary site for a particular piece of information and scale to the amount of information that exists, not only between all computers in a planetary-scale distributed system, but all entities in the universe?

\section*{Acknowledgements}
Thanks to John Muellerleile for helping me initially develop this idea and Colin Barrett, Michael R. Bernstein, Paul Borrill, Sean Cribbs, Chas Emerick, Scott Lystig Fritchie, Caitie McCaffrey, Tom Santero, Justin Sheehy, and Jordan West for their feedback.

\bibliographystyle{abbrv}
\bibliography{tendency-article}
\end{document}